\documentclass[twocolumn,nofootinbib,preprintnumbers,superscriptaddress,amsmath,amssymb,PRL]{revtex4-1}
\usepackage{graphicx}
\usepackage{dcolumn}
\usepackage{float}
\usepackage{mathptmx, courier, pifont}
\usepackage[scaled=0.92]{helvet}
\usepackage[T1]{fontenc}
\usepackage{textcomp}
\usepackage{color}
\usepackage[normalem]{ulem}

\usepackage[dvipsnames]{xcolor}
\usepackage[colorlinks=true,urlcolor=Blue,linkcolor=Blue]{hyperref}
\usepackage[all]{hypcap}
\usepackage{makecell}
\usepackage{footmisc}
\usepackage{bibunits}
\newcommand{\sortas}[1]{}
\begin{document}
%\begin{bibunit}

%\linenumbers 

%%%%%%%%%%%%%%%%%%%%%%%%%%
%
% TITLE: MAX 15 WORDS
%
%%%%%%%%%%%%%%%%%%%%%%%%%%

\title{$^{159}$Dy electron-capture: new candidate for neutrino mass determination}

\author{Z.~Ge}\thanks{Corresponding author: zhuang.z.ge@jyu.fi}
\affiliation{Department of Physics, University of Jyv\"askyl\"a, P.O. Box 35, FI-40014, Jyv\"askyl\"a, Finland}%
\author{T.~Eronen}\thanks{Corresponding author: tommi.eronen@jyu.fi}
\affiliation{Department of Physics, University of Jyv\"askyl\"a, P.O. Box 35, FI-40014, Jyv\"askyl\"a, Finland}%
\author{K.~S.~Tyrin}
\affiliation{National Research Centre ``Kurchatov Institute'', Ploschad' Akademika Kurchatova 1, 123182 Moscow, Russia}%
%Kotila, Jenni-Mari 
\author{J.~Kotila}
\affiliation{Finnish Institute for Educational Research, University of Jyv\"askyl\"a, P.O. Box 35, FI-40014, Jyv\"askyl\"a, Finland}%
\affiliation{Center for Theoretical Physics, Sloane Physics Laboratory Yale University, New Haven, Connecticut 06520-8120, USA}%
\author{J.~Kostensalo}
\affiliation{Department of Physics, University of Jyv\"askyl\"a, P.O. Box 35, FI-40014, Jyv\"askyl\"a, Finland}%
\author{D.~A.~Nesterenko}
\affiliation{Department of Physics, University of Jyv\"askyl\"a, P.O. Box 35, FI-40014, Jyv\"askyl\"a, Finland}%
\author{O.~Beliuskina}
\affiliation{Department of Physics, University of Jyv\"askyl\"a, P.O. Box 35, FI-40014, Jyv\"askyl\"a, Finland}%
\author{R.~de~Groote}
\affiliation{Department of Physics, University of Jyv\"askyl\"a, P.O. Box 35, FI-40014, Jyv\"askyl\"a, Finland}%
\author{A.~de~Roubin}
\affiliation{Centre d'Etudes Nucl\'eaires de Bordeaux Gradignan, UMR 5797 CNRS/IN2P3 - Universit\'e de Bordeaux, 19 Chemin du Solarium, CS 10120, F-33175 Gradignan Cedex, France}%
\author{S.~Geldhof}\thanks{Present address: KU Leuven, Instituut voor Kern- en Stralingsfysica, B-3001 Leuven, Belgium}
\affiliation{Department of Physics, University of Jyv\"askyl\"a, P.O. Box 35, FI-40014, Jyv\"askyl\"a, Finland}%
%\altaffiliation[]{Present address: KU Leuven, Instituut voor Kern- en Stralingsfysica, B-3001 Leuven, Belgium}
%\alsoaffiliation{lab 2}  \altaffiliation{}
%\altaffiliation{Current address: Institut2 woanders}
%\affiliation[Institut1]{Institut1, Stadt, Land}
\author{W.~Gins}
\affiliation{Department of Physics, University of Jyv\"askyl\"a, P.O. Box 35, FI-40014, Jyv\"askyl\"a, Finland}%
\author{M.~Hukkanen}
\affiliation{Department of Physics, University of Jyv\"askyl\"a, P.O. Box 35, FI-40014, Jyv\"askyl\"a, Finland}%
%\affiliation{University of Bordeaux,....?}
%\affiliation{Centre d\’Etudes Nucl\´eaires de Bordeaux Gradignan, UMR 5797 CNRS/IN2P3 - Universit\´e de Bordeaux, 19 Chemin du Solarium, CS 10120, F-33175 Gradignan Cedex, France}
\affiliation{Centre d'Etudes Nucl\'eaires de Bordeaux Gradignan, UMR 5797 CNRS/IN2P3 - Universit\'e de Bordeaux, 19 Chemin du Solarium, CS 10120, F-33175 Gradignan Cedex, France}
\author{A.~Jokinen} 
\affiliation{Department of Physics, University of Jyv\"askyl\"a, P.O. Box 35, FI-40014, Jyv\"askyl\"a, Finland}%
\author{A.~Kankainen}
\affiliation{Department of Physics, University of Jyv\"askyl\"a, P.O. Box 35, FI-40014, Jyv\"askyl\"a, Finland}%
\author{\'A.~Koszor\'us}
\affiliation{Department of Physics, University of Liverpool, Liverpool, L69 7ZE,  United Kingdom}%
\author{M.~I.~Krivoruchenko}\thanks{Corresponding author: mikhail.krivoruchenko@itep.ru}
\affiliation{National Research Centre ``Kurchatov Institute'', Ploschad' Akademika
Kurchatova 1, 123182 Moscow, Russia}%
\affiliation{Institute for Theoretical and Experimental Physics, NRC ``Kurchatov
Institute'', B. Cheremushkinskaya 25, 117218 Moscow, Russia}
\author{S.~Kujanp\"a\"a}
\affiliation{Department of Physics, University of Jyv\"askyl\"a, P.O. Box 35, FI-40014, Jyv\"askyl\"a, Finland}%
\author{I.~D.~Moore}
\affiliation{Department of Physics, University of Jyv\"askyl\"a, P.O. Box 35, FI-40014, Jyv\"askyl\"a, Finland}%
\author{A.~Raggio}
\affiliation{Department of Physics, University of Jyv\"askyl\"a, P.O. Box 35, FI-40014, Jyv\"askyl\"a, Finland}%
\author{S.~Rinta-Antila}
\affiliation{Department of Physics, University of Jyv\"askyl\"a, P.O. Box 35, FI-40014, Jyv\"askyl\"a, Finland}%
\author{J.~Suhonen}
\affiliation{Department of Physics, University of Jyv\"askyl\"a, P.O. Box 35, FI-40014, Jyv\"askyl\"a, Finland}%
\author{V.~Virtanen}
\affiliation{Department of Physics, University of Jyv\"askyl\"a, P.O. Box 35, FI-40014, Jyv\"askyl\"a, Finland}%
%\author{I.~D.~Moore} 
%Andrew  Weaver
%Sasha 	Zadvornaya
%Andrea Raggio
\author{A.~P.~Weaver}
\affiliation{School of Computing, Engineering and Mathematics, University of Brighton, Brighton BN2 4JG, United Kingdom}%
\author{A.~Zadvornaya}
\affiliation{Department of Physics, University of Jyv\"askyl\"a, P.O. Box 35, FI-40014, Jyv\"askyl\"a, Finland}%
%\pacs{21.10.Dr, 27.40.+z, 29.20.db}% 26.30.Ca,
%\pacs{21.10.Dr, 27.40.+z, 29.20.db}% 26.30.Ca,  Atomic Mass  Evaluation 2020 (AME2020) $Q$ value
\date{\today}

\begin{abstract}
{
The ground-state to ground-state  electron-capture $Q$ value of $^{159}$Dy ($3/2^-$) has been measured directly utilizing the double Penning trap mass spectrometer JYFLTRAP. A value of 364.73(19)~keV was obtained from a measurement of the cyclotron frequency ratio of the decay parent $^{159}$Dy and the decay daughter $^{159}$Tb ions
using the novel phase-imaging ion-cyclotron resonance technique.  The $Q$ values for allowed Gamow-Teller transition to $5/2^-$ and the third-forbidden unique transition to $11/2^+$ state
with excitation energies of 363.5449(14)~keV and 362.050(40)~keV in $^{159}$Tb were determined to be  1.18(19) keV and  2.68(19) keV, respectively. The high-precision $Q$ value of transition $3/2^-\to 5/2^-$ from this work, revealing itself as the lowest electron-capture $Q$ value, is utilized to unambiguously characterise all the possible lines that are present in its electron capture spectrum. 
%Based on the experimental results, 
%\textcolor{red}
{
We performed atomic many-body calculations for both transitions to determine electron-capture probabilities from various atomic orbitals, and found an order of magnitude enhancement in the event rates near the end-point of energy spectrum in the transition to the $5/2^-$ nuclear excited state,
%compared to $^{163}$Ho, 
which can become very interesting once the experimental challenges of identifying decays into excited states are overcome. The transition to the $11/2^+$ state is strongly suppressed and found unsuitable for measuring the neutrino mass. 
%These results show that the electron capture in the $^{159}$Dy atom, going to the $5/2^-$ state of the $^{159}$Tb nucleus, is a process that opens the way to long-term and high sensitive experiments of a new generation for direct determination of neutrino mass.
%up to sub-eV sensitivity.
These results show that the electron capture in the $^{159}$Dy atom, going to the $5/2^-$ state of the $^{159}$Tb nucleus, %\textcolor{red}
{is a new candidate which may open the way to determine the electron-neutrino mass in the sub-eV region by studying EC. Further experimental feasibility studies, including coincidence measurements with realistic detectors, will be of great interest.}
}
%We performed atomic many-body calculations for both transitions to determine the electron-capture rates from different atomic shells and found an order-of-magnitude enhancement in the event rate near the endpoint for the transition to $5/2^-$ state compared to $^{163}$Ho. The  EC  transition  to the $11/2^+$  state was found to be not potential for neutrino-mass measurements.
%\textbf
%These findings reveal that $^{159}$Dy with transition to $5/2^-$ state in $^{159}$Dy is a suitable ground-to-excited state decay candidate to pursue a sub-eV sensitivity on the electron neutrino mass opening up ways towards long-term and high-sensitivity new-generation experiments for direct neutrino mass determination.
%\textcolor{blue}{
%These findings reveal that $^{159}$Dy with transition to $5/2^-$ state in $^{159}$Tb is a suitable ground-to-excited state decay candidate opening up ways towards long-term and high-sensitivity new-generation experiments for direct neutrino mass determination down to sub-eV sensitivity.
%}
}
\end{abstract}
\maketitle

%\clearpage
%\newpage

The neutrino is perhaps the most mysterious particle of all elementary particles.  The problem of the overall scale of neutrino masses is a matter of paramount importance in the search for generalizations of the Standard Model, as well as for cosmology. 
%The experimental discovery of neutrino oscillations by the Super-Kamiokande Collaboration~\cite{Fukuda1998} showed that neutrinos have non-zero masses. 
%Numerous modern experiments on neutrino %oscillations~\cite{Fukuda1998,Bilenky:2019,Gonzalez-Garcia:2020} allow to extract %non-zero differences in the squared neutrino masses and mixing angles, but not the %absolute values.
Numerous modern experiments on neutrino oscillations~\cite{Fukuda1998,Bilenky:2019,Gonzalez-Garcia:2020} allow extracting non-zero
differences between the neutrino masses-squared and also the oscillation parameters.
These experiments are insensitive to the overall scale of neutrino masses. However,
they limit the effective mass of electron antineutrino to be at least 0.048~eV/$c^2$ and 0.0085~eV$/c^2$ for the inverted and normal mass orderings, respectively~\cite{Gonzalez-Garcia:2020}. 
%, but do not say anything about the absolute values of the masses. establish
%From neutrino oscillations, only the differences of the masses squared can be extracted, but not the absolute values.
%The fact that neutrinos are massive is the strongest demonstration that the Standard Model is incomplete and there must be new physics beyond the Standard Model. 
Indirect measurements of the neutrino mass, which 
also allow to clarify the Dirac or Majorana nature of neutrinos, are conducted in the search of neutrinoless double-$\beta^-$~decay 
%\textcolor{red}
{with a sensitivity of about 0.1~eV/$c^2$}~\cite{Suhonen1998,Avignone2008,Vergados:2012xy,Ejiri2019} and neutrinoless double-electron capture~\cite{Blaum2020}. 
The only direct and model-independent methods for measuring the mass of neutrinos are based on the study of single-electron capture (EC)~\cite{Gastaldo2017}, while the mass of antineutrinos is measured in single-$\beta^-$ decays~\cite{aker2021direct}.

Presently, the most stringent upper limit of 0.8~eV/c$^2$ (90\% Confidence Level (C.L.))  for the effective electron anti-neutrino mass $m_{\overline{\nu}_e}$ originates from very recent data obtained with the KATRIN (KArlsruhe TRitium Neutrino) experiment, by investigating the emitted electron spectrum endpoint of tritium $\beta^-$ decay~\cite{aker2021direct}. 
The most stringent upper limit of the effective electron neutrino mass $m_{\nu_e}$ is as large as 150~eV/c$^2$~\cite{Velte2019}  (95\% C.L.), derived from the analysis of the EC endpoint of $^{163}$Ho, which is being utilized for next-generation direct neutrino-mass determination experiments such as  ECHo~\cite{Gastaldo2017} and HOLMES~\cite{Faverzani2016}.

%
%The value has come down from 225 eV/c$^2$~\cite{Springer1987} in the past 40~years. Evidently these experiments are extremely challenging, in the most part due to the very small fraction of decay events ($\sim$10$^{-12}$ for $^{163}$Ho) that fall into the most relevant $\approx 1$~eV energy below the endpoint. 
%The difference in $|m_{\overline{\nu}_e}|$ and $|m_{\nu_e}|$
%the absolute values of the effective masses of the electron anti-neutrino and the electron neutrino 
%will indicate a violation of the charge conjugation -- parity -- time reversal (CPT) symmetry. 

%-----------------------------Table 1 +2--------------------------------
%%%%%%%%%%%%%%%%%%%%%%% ~\cite{Larkins1977}
%{Extended Data Table 1 | $Q$ values of the EC transitions from the $3/2^-$ ground state (gs) of the parent nucleus $^{159}$Dy to the potential low-$Q$-value excited states of the daughter nucleus $^{159}$Tb adopted from literature.}
\begin{table*}[!htbp]
   \caption[]{$Q$ values of the EC transitions from the $3/2^-$ ground state (gs) of the parent nucleus $^{159}$Dy to the potential low-$Q$-value excited states of the daughter nucleus $^{159}$Tb.
   %adopted from literature or this work. 
   %\textbf{ | $Q$ values of the EC transitions from literature.}  Summary of the transitions from the $3/2^-$ ground state (gs) of the parent nucleus $^{159}$Dy to the potential low-$Q$-value excited states of the daughter nucleus $^{159}$Tb adopted from literature. 
   The first column indicates the excited final state of interest for the low  $Q$-value transition or the ground state with spin-parities indicated. The second column gives the decay type.  The third column lists the experimental excitation energy $E^{*}_i$ with the experimental error~\cite{NNDC} and the fourth column gives the derived experimental decay $Q_{\mathrm{EC}}^{i}$ ($i$ = 1, 2) value. %Further 
   The fifth to ninth  columns denoted as
   %total energy 
  % $Q^{i}_{x}
  $\Delta^{i}_{x}$
  %, = Q^{i}_{EC} - \varepsilon_x$, 
  % of the emitted neutrino in the possible allowed shell-capture (M1, M2, N1, N2, O1) decay. 
  give the distance of the $Q_{\mathrm{EC}}^{i}$  value to the binding energy $\varepsilon_x$ (from~\cite{X-Ray_Data_Booklet}) of the electrons in the daughter atoms.
  %Negative values indicate that the capture is prohibited while positive values are for allowed captures.
  The last  column indicates the source of values on each raw,  either from Atomic mass evaluation literature (AME2020)~\cite{Huang2021,Wang2021} or obtained from this work.
   %$Q^{i}_{EC} = Q^{gs}_{EC} - E^{*}_i$
   %%:  $Q^{i}_{\nu,x} = Q^{i}_{EC} - e_{x}$.
   %\footref{fn:a}.
   %\footnotemark[\ref{fn:a}].
   %\ref{fn:a}
   %\footref[\ref{fn:a}].
   %\footnote{\label{a}Captures of electrons occupying the K and L shells are energetically forbidden. Only electrons from $s$ and $p_{1/2}$-levels from the third and higher shells (M1, M2, N1, N2, O1, O2, P1, \ldots) can possibly be captured due to angular momentum conservation and the finite overlap of their wave function with the nucleus.}.
   %from this work
   %The atomic binding energies $\varepsilon_x$ of the daughter are adopted from~\cite{X-Ray_Data_Booklet}. 
   The decay $Q$ values and excitation energies  are in units of keV.
   %Here 3rd FU means 3rd forbidden unique.
   }
  \begin{ruledtabular}
   \begin{tabular*}{\textwidth}{cccccccccc}
   %   \hline
   %   \hline
   Final state &Decay type & $E_i^{*}$ & 
   $Q_{\mathrm{EC}}^{i}$   
   %   $Q_{\mathrm{EC}}^{i}$   
   &$\Delta^{i}_{\mathrm{M1}}$
 %     &$Q^{i}_{\mathrm{M1}}$
   %&\makecell[c]{ $Q^{i}_{\mathrm{M1}}$\\ ({$\varepsilon_\mathrm{M1}$}: 1.968)}
   &$\Delta^{i}_{\mathrm{M2}}$
%  & \makecell[c]{$ Q^{i}_{\mathrm{M2}}$ \\ ({$\varepsilon_\mathrm{M2}$}:  1.768)}&\makecell[c]{$ Q^{i}_{\mathrm{N1}}$ \\ ({$\varepsilon_\mathrm{N1}$}:  0.396)}
   &$\Delta^{i}_{\mathrm{N1}}$
   %\makecell[c]{$ Q^{i}_{\mathrm{N2}}$ \\ ({$\varepsilon_\mathrm{N2}$}:  0.3224)} 
   &$\Delta^{i}_{\mathrm{N2}}$
   %\makecell[c]{$ Q^{i}_{\mathrm{O1}}$ \\ ({$\varepsilon_\mathrm{O1}$}:  0.0456)} \\
   %$Q_{\mathrm{EC}}^{i}$   
   & $\Delta^{i}_{\mathrm{O1}}$
   &Source\\
   %%%%%\makecell[c]{ $Q^{i}_{\nu,\mathrm{M1}}$\\ ({$\varepsilon_\mathrm{M1}$}: 1.968)} &\makecell[c]{$ Q^{i}_{\nu,\mathrm{M2}}$ \\ ({$\varepsilon_\mathrm{M2}$}:  1.768)}&\makecell[c]{$ Q^{i}_{\nu,\mathrm{N1}}$ \\ ({$\varepsilon_\mathrm{N1}$}:  0.396)}&\makecell[c]{$ Q^{i}_{\nu,\mathrm{N2}}$ \\ ({$\varepsilon_\mathrm{N2}$}:  0.3224)} &\makecell[c]{$ Q^{i}_{\nu,\mathrm{O1}}$ \\ ({$\varepsilon_\mathrm{O1}$}:  0.0456)} \\
\hline\noalign{\smallskip}
 5/2$^-$ & {Allowed} & 363.5449(14) & \makecell[c]{ 1.7(12) \\1.18(19)}& \makecell[c]{-0.3(12) \\-0.78(19) }&\makecell[c]{-0.1(12) \\-0.58(19)}&\makecell[c]{ 1.3(12) \\0.79(19)}&\makecell[c]{ 1.3(12) \\0.87(19)} & \makecell[c]{1.6(12) \\1.14(19)}&\makecell[c]{ AME2020\\ this work}\\  \hline
 11/2$^+$  &   \makecell[c]{3rd forbidden\\unique}&362.050(40) &  \makecell[c]{3.2(12) \\2.68(19)  }&\makecell[c]{1.2(12) \\ 0.63(19) }&\makecell[c]{1.4(12)\\0.84(19) }&\makecell[c]{ 2.8(12)\\ 2.26(19) }&\makecell[c]{2.8(12)\\2.35(19) }&\makecell[c]{3.1(12)\\2.62(19) }&\makecell[c]{ AME2020\\ this work} \\ \hline
3/2$^{+}$&  & 0&\makecell[c]{  365.2(12) \\364.73(19) } && & &&&\makecell[c]{ AME2020\\ this work}\\
   \end{tabular*}
   \label{table:Q-value-AME}
   \end{ruledtabular}
\end{table*}
%%%%%%%%%%%%%%%%%%%%%%%970.83(64)
%-----------------------------Table 1 +2--------------------------------

The search for potential isotopes for possible future long-term and high-sensitivity (anti)neutrino-mass determination experiments~\cite{Ejiri2019,Mustonen2010,Suhonen2014,DeRoubin2020,ge2021} in the pursuit of sub-eV sensitivity, is of great interest. 
% modifed
%A potential EC case preferably has its $Q$ value close to one of the atomic binding energies of the captured electron. This enhances the decay event rate near the endpoint, where the effects of a non-vanishing neutrino mass are relevant. 
%For $\beta^-$ decay spectra, the event rate dependence on the $Q$ value near the endpoint scales with $Q^{-3}$, but for EC instead it is steeper and, potentially, radically enhanced.
%
%
%For EC, the closer is the $Q$ value of the decay to one of the ionization energies of the captured electrons, the larger the resonance enhancement of the rate near the endpoint, where the effects of a non-vanishing anti-neutrino mass are relevant. The event rate dependence on $Q$ value near the endpoint for EC is deeper than that of $\beta^-$ decay. 
%
%referee asked to explain in more details
For $\beta^-$ decay spectra, the neutrino mass sensitivity depends on the fraction of events close to the endpoint, where the cumulative decay rate is proportional to the phase-space factor and scales with $Q^{-3}$.
%
%For EC, the event rate dependency on $Q$ value near the endpoint is more sensitive and depends on how close the decay $Q$ value is to the ionization energy of the captured electron.
%
%\textcolor{red}
{
For EC, the cumulative event rate near the endpoint is proportional to $Q^{-2}$, and it increases when the electron orbitals have an ionization energy close to the value of $Q$.
Nuclides, favored for such direct neutrino mass experiments, are the ones with a small $Q$ and the electron orbitals close to the threshold.
}
%Nuclides, favored for such direct neutrino mass experiments, are the ones with a small $Q$ value.
% The following 
%The enhancement of the event rate is notably strong for $^{159}$Dy ($t_{1/2} = 144.4(2)$~days, $J^\pi = 3/2^-$), which makes it a particularly interesting and a possible competitor to $^{163}$Ho, the latter having a $Q$ value of 2.833(30)$_{\rm stat}$(15)$_{\rm sys}$ keV~\cite{Eliseev2015}. $^{163}$Ho is being utilized for several next-generation direct neutrino-mass determination experiments such as  ECHo (Electron Capture in $^{163}$Ho)~\cite{Gastaldo2017} and HOLMES~\cite{Faverzani2016}.
%, and NuMECS~\cite{Croce2016}. 
$^{159}$Dy, studied here, decays only by EC and its ground-to-ground state $Q$ value ($Q_{\mathrm{EC}}^{\mathrm{gs}}$) 365.2(12)~keV~\cite{Huang2021,Wang2021} is close to the excitation energies ($E^{*}_i$)~\cite{NNDC} of two candidate excited states having spin-parity $5/2^-$ and $11/2^+$ in the daughter nucleus $^{159}$Tb, see Table~\ref{table:Q-value-AME}. The EC $Q$ values to the excited states 
%, $Q^{i}_{EC} = Q^{gs}_{EC} - E^{*}_i$,
are expected to be very small. Especially EC to the $5/2^-$ state is of significant interest since it is of Gamow-Teller type and has been experimentally confirmed to exist with a branching ratio $1.9(5)\times10^{-6}$~\cite{NSR1968MY01}.
%!!!!added as suggested by referee
Branching ratio of EC to the $11/2^+$ state is tiny compared to $5/2^-$ state and this decay branch has not been observed.
%!!!
The total energy of the neutrino emitted in EC decay is determined by the atomic binding energies of the possible allowed atomic shells of the captured electron. In the present case, captures of electrons occupying the K and L shells for the transition $^{159}$Dy(3/2$^-$) $\rightarrow$ $^{159}$Tb$^*$(5/2$^-$) are energetically forbidden. Only electrons from $s$ and $p_{1/2}$-levels from the third and higher shells (M1, M2, N1, N2, O1, O2, and P1) can possibly be captured due to angular momentum conservation and the finite overlap of their wave function with the nucleus. 
%:  $Q^{i}_{\nu,x} = Q^{i}_{EC} - e_{x}$.
This makes the EC energy even smaller, as tabulated in Table~\ref{table:Q-value-AME}.
The nuclear excitation energies of the two daughter states are already rather accurately known ($< 40$~eV). The main uncertainty in the $Q$ value is due to the 1.2~keV uncertainty in the ground-to-ground state $Q$ value, which is primarily determined from $^{159}$Dy(EC)$^{159}$Tb decay data~\cite{Huang2021,NSR1968MY01,NNDC}. 
With this large uncertainty it is impossible to model the EC spectrum shape, especially near the endpoint where the decay rate is extremely sensitive to the $Q$ value. The current precision does not even allow an order-of-magnitude scale estimate.

In this letter, we report on the first direct $^{159}$Dy ground-to-ground state EC $Q$-value determination.
%via cyclotron frequency measurements with high-precision Penning trap mass spectrometer.
Based on the results, we performed atomic many-body calculations in order to determine the partial EC rates from different atomic shells for the two discussed EC transitions: the allowed Gamow-Teller transition $3/2^-\to 5/2^-$ and the third-forbidden unique transition $3/2^-\to 11/2^+$. We have also determined the partial half-lives of the captures from different atomic shells for the Gamow-Teller transition by normalizing to the measured total EC branching to the $5/2^-$ state. 
The measurements were conducted at the Ion Guide Isotope Separator On-Line facility (IGISOL) using the double Penning trap mass spectrometer JYFLTRAP~\cite{Eronen2012} in the accelerator laboratory of University of Jyv\"askyl\"a, Finland~\cite{Moore2013}. 
%Penning trap mass spectrometry allows for the determination of the $^{159}$Dy $Q$ value directly through the cyclotron frequency ratio measurement of the decay-parent $^{159}$Dy and decay-daughter $^{159}$Tb ions.
%
%The layout of the experimental facility and its relevant parts is visualized in Fig.~\ref{fig:igisol}. %Fig.~\ref{fig:setup_and_experimental_data_visualized} (a). 
%A comprehensive description can be found from Refs.~\cite{Eronen2012a,Moore2013}.
%
%
To produce $^{159}$Dy$^+$ ions, a proton beam of 40 MeV in energy from the K-130 cyclotron was used to bombard dysprosium target with natural abundance. % to produce the ions of $^{159}$Dy$^+$.
Ions of stable daughter $^{159}$Tb$^+$ were separately produced with an offline glow-discharge ion source.
%Irrespective of the source, the ion beam was transported through a dipole magnet with a mass resolving power $\approx$ 500, which was sufficient to separate the $A=159$ isobar. The selected ions were transported into a radiofrequency quadrupole cooler-buncher (RFQ)~\cite{Nieminen2001} and further to the JYFLTRAP double Penning trap, consisting of two cylindrical Penning traps which are both situated inside a 7-T superconducting solenoid.
%
%The $^{159}$Dy$^+$ ions were accompanied with co-produced unwanted contaminants $^{159}$Ho$^+$, $^{159m}$Ho$^+$ and, to a lesser extent, stable $^{159}$Tb ions. It was necessary to remove these ions prior to the frequency measurement to avoid ion-ion related frequency shifts. The states of $^{159}$Ho$^+$ were removed with the buffer gas cooling method~\cite{Savard1991} in the preparation trap while $^{159}$Tb$^+$ ions were removed with dipolar excitation in the precision trap~\cite{Eronen2008a}. For the  offline-produced $^{159}$Tb ions no additional purification was needed.

The phase-imaging ion-cyclotron resonance (PI-ICR) technique~\cite{Eliseev2014,Nesterenko2018} 
was used to measure the cyclotron frequencies $\nu_{c}=\frac{1}{2\pi}\frac{q}{m}B$, where $q/m$ is the charge-to-mass ratio of the measured $^{159}$Dy$^+$ and $^{159}$Tb$^+$ ions and $B$ the magnetic field. We used the scheme that allows direct determination of $\nu_c$ via the sideband coupling frequency $\nu_c = \nu_+ + \nu_-$, where $\nu_+$ is the trap-modified cyclotron frequency and $\nu_-$ the magnetron frequency. %~\cite{Blaum2006}. 
Phase accumulation time $t = 514$~ms was chosen for both $^{159}$Dy$^+$ and $^{159}$Tb$^+$ ions to ensure that the spot of interest was resolved from any leaked isobaric, isomeric and molecular contamination. No contaminating ions were observed.
\begin{figure}[!htp]
   %\flushleft
   \includegraphics[width=0.80\columnwidth]{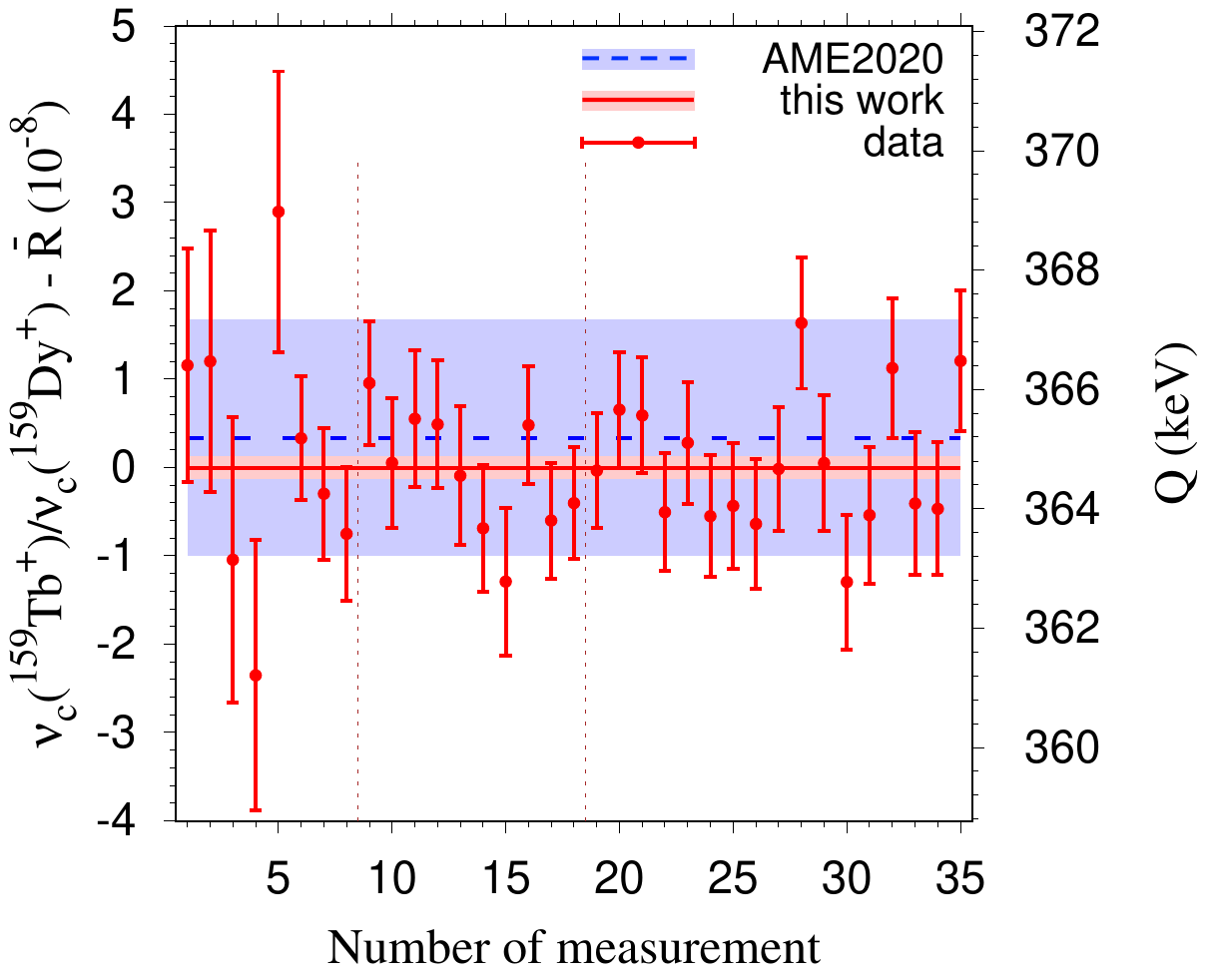}
   \caption{(Color online). Comparison of results obtained in this work and the %Atomic mass evaluation 
   literature value (AME2020)~\cite{Huang2021,Wang2021}. In total 36 individual frequency ratios were measured in 3 time slots.
   %and used to obtain the final value. 
   %The final weighted mean frequency ratio $\overline{R}$ is 1.000 002 463 8(13). The $Q_{\mathrm{EC}}^{\mathrm{gs}}$ from this work, 364.73(19)~keV, is a factor of 6.3 times more precise and 0.47 keV smaller than the literature value.
   }
   \label{fig:data-points}
\end{figure}

%-----------------------------Fig. 2 --------------------------------$B_\textrm{e}$
%\begin{figure}[!htp]
   %\flushleft
%   \includegraphics[width=0.80\columnwidth]{Fig2_cyc-mag-center-2panel.pdf}
   %{Fig2.pdf}{Fig2-data.pdf}
%   \caption{(Color online). 
%   Ion spots (center, cyclotron phase and magnetron phase) of $^{159}$Dy$^{+}$ on the 2-dimensional position-sensitive MCP detector. %after a typical PI-ICR excitation pattern with an accumulation time of 514 ms. 
   %The magnetron spot is displayed on the left side and the cyclotron phase phase spot on the right. The angle difference between the two spots relative to the center spot is utilized to deduce the cyclotron frequency of the measured ion species.  
%   The color bar illustrates the number of ions in each pixel.
   %Comparison of results obtained in this work and the Atomic mass evaluation literature value (AME2020)~\cite{Huang2021,Wang2021}. In total 36 individual frequency ratios were measured and used to obtain the final value. The final weighted mean frequency ratio $\overline{R}$ is 1.000 002 463 8(13). The $Q_{\mathrm{EC}}^{\mathrm{gs}}$ from this work, 364.73(19)~keV, is a factor of 6.3 times more precise and 0.47 keV smaller than the literature value.
%   }
%    \label{fig:2-phases} 
%\end{figure}

%\textbf{Experimental analysis}
%\textbf{$Q$-value determination of $^{159}$Dy-$^{159}$Tb.}
The parent $^{159}$Dy and daughter $^{159}$Tb ion cyclotron frequency measurements were interleaved by changing between the two every 3 minutes to minimize the uncertainty contribution of the magnetic field fluctuation in the measured cyclotron frequency ratio. 
%An interpolation method was used to temporally match the measurements of the two species. In total, 8.5 hours of frequency ratio data was collected.
The data was analyzed by dividing the collected data to approximately 12 minute sections in order to have reasonable amount of statistics for fitting the magnetron and cyclotron phase spots using the maximum likelihood method. Ion bunches up to 5 detected ions were used in the analysis. Additionally, a countrate-class analysis~\cite{Kellerbauer2003}, in which the frequency data were split up by the number of ions simultaneously present in the precision trap, was carried out. No evidence of a correlation between frequency measurements and count rate was observed, which indicated no systematic frequency shifts at the achieved statistical precision level due to ion-ion interactions. 
Furthermore, $^{159}$Dy and $^{159}$Tb ions being mass doublets cancel many of the systematic uncertainties in the cyclotron frequency ratio~\cite{Roux2013}.

The $Q_{\mathrm{EC}}^{\mathrm{gs}}$ is obtained from the mass difference of $^{159}$Dy and $^{159}$Tb utilizing the mass-energy equivalence formula $E=mc^2$:
 \begin{equation}
\label{eq:Qec}
Q_{\mathrm{EC}}=(M_{\mathrm{i}} - M_{\mathrm{f}})c^2 = (R-1)(M_{\mathrm{f}} - m_{\mathrm{e}})c^2+\Delta B_{\mathrm{if}},
% m=r(m_{ref}-m_e)+m_e,Q_{\mathrm{EC}}
 \end{equation}
where $M_{\mathrm{i}}$ and $M_{\mathrm{f}}$ are the atomic masses of the parent and daughter atoms, respectively, and $R = {\nu_{c,\mathrm{f}}}/{\nu_{c,\mathrm{i}}}$ is their cyclotron frequency ratio obtained in charge state 1+. The value $\Delta B_{\mathrm{if}}$ describes the contribution from electron binding energy differences of the parent and daughter atoms (here 0.07525(60)~eV for $^{159}$Dy$^+$ and $^{159}$Tb$^+$~\cite{NIST_ASD}). $m_{\mathrm{e}}$ is the mass of electron.
%The weighted mean ratio $\overline{R}$ of the single ratios for PI-ICR data was calculated along with the inner and outer errors~\cite{Birge1932}.  The maximum of the inner and outer errors of the ratios were taken as the weights to calculate $\overline{R}$. 
%The ${Q}^{\mathrm{gs}}_{\mathrm{EC}}$ value was derived based on Eq. (\ref{eq:Qec}). The individual frequency ratios and derived $Q^{gs}_{EC}$ values from this work are shown in Fig.~\ref{fig:data-points}.
%
%
%Three sets of data were collected (see Fig.~\ref{fig:data-points}) and the final weighted mean ratio $\overline{R}$ was calculated from their averages.
%If normalized $\chi^2$ was more than one, the uncertainty of the set was expanded with square root of it~\cite{Birge1932}.
%
%\textcolor{red}
{Three sets of data were collected (see Fig.~\ref{fig:data-points}). 
The normalized $\chi^2$ for the sets were 1.2, 0.9 and 1.0. The uncertainty of the first was expanded with the square-root of it, marginally affecting the final
weighted mean ratio $\overline{R}$.}
The final weighted mean frequency ratio $\overline{R}$ is 1.000 002 463 8(13), which results in ${Q}^{\mathrm{gs}}_{\mathrm{EC}}$  = 364.73(19)~keV.

The obtained $Q_{\mathrm{EC}}^{\mathrm{gs}}$ from this work  is more than six times more precise and 0.47 keV smaller than the AME2020 value, which was derived primarily from an EC decay measurement of $^{159}$Dy(EC)$^{159}$Tb~\cite{Huang2021}.
%
%Using the newly determined  $Q_{\mathrm{EC}}^{\mathrm{gs}}$, the possibility of EC to known excited states in $^{159}$Tb is investigated. 
The newly measured high-precision $Q_{\mathrm{EC}}^{\mathrm{gs}}$, together with the accurate nuclear energy level data, yields $Q_{\mathrm{EC}}^i$ values of 1.18(19)~keV and 2.68(19)~keV for the $5/2^-$ and $11/2^+$ states in $^{159}$Tb,  respectively.
%{table:Q-value}).
$Q$ values of different atomic electron shell captures are  tabulated in Table~\ref{table:Q-value-AME}. %Table~\ref{table:Q-value}. 
%A comparison of the $Q$ values of the ground-to-excited-state EC transitions from this work to the values derived from AME2020 are shown in Fig.~\ref{fig:level-sheme}. 
Which orbital electrons take part in the EC process and the absolute $Q$ values of the decays are crucial for modelling the spectrum shape near the endpoint.
%Our results reveal that EC to the $5/2^-$ state can occur only from the N1 orbital or higher. 
%Move the following to the caption of the figure
In this work, M2 capture to the $5/2^-$ state is confirmed to be energetically forbidden at 3.3$\sigma$ level, revealing N1 to be the first energetically possible capture at $4.0\sigma$ level.
In addition, the M1 capture to the $11/2^+$ state is confirmed to be positive at 3.7$\sigma$ level and captures can proceed from M1 and higher orbits.
The unambiguous characterization of all the possible lines in the EC spectrum at a significance level of at least 3$\sigma$ for the transitions, makes the modelling of their shape possible.
%For captures to the $11/2^+$ state, they can proceed from M1 and higher orbits.
%In addition, the M1 capture to the $11/2^+$ state is confirmed to be positive at 3.7$\sigma$ level.
% Thus, the EC spectra of the decays to the two excited states are definitely confirmed and characterized.  
%These are crucial findings for modelling the spectrum shape of the EC to $5/2^-$ state. The first possibility is the N1 capture while M1, M2 and further inner shell captures have now been ruled out at least with 3.3$\sigma$ level. 
%%%
%
%clearly indicate from which atomic shells the EC can proceed. This is crucial data for modelling the EC spectrum shape for neutrino mass determination.

%   THEORY

To estimate the EC partial half-lives and the distribution of energy released in the decays, we have performed Dirac-Hartree-Fock atomic many-body calculations.
% (see Methods).
%
The EC capture rate is determined by the standard $\beta$-decay Hamiltonian. The probability depends on the wave function of the electrons inside the nucleus, on the exchange-and-overlap factor of the spectator electrons due to the non-orthogonality of the atomic shells of the parent and daughter atoms, as well as the nuclear matrix element.
% Some details of the calculations are given in the section "Methods".

%-----------------------------Table 3+4 -------------------------------- Extended Data 
%%%%%%%%%%%%%%%%%%%%%%%
%   Resulting $\\lambda$ for capture from individual orbitals in units of $\hslash$ = $c$ = $m_e$ = 1. Experimental binding energies for daughter Tb atom are taken from X-RAY DATA BOOKLET[1]. (1)-Binding energy calculated in GRASP for isolated atom with configuration [Xe]4f$^{10}$6s$^{1}$.
%measured branching to the 5/2- state, giving a partial half-life of  2.08 $\times 10^5$  years. You can then normalize the various EC partial half-lives to this total partial half-life. These partial half-lives
  %EC (5/2$^-$ excited state) half-lives $T_{1/2}$ of the various levels for $g_\mathrm{A}$ = 1.
\begin{table}[htbp!]
   \caption
   %[\textbf{ | Normalized partial half-lives for the Gamow-Teller EC transition $3/2^- \to 5/2^-$.}]
   {Normalized partial half-lives for the Gamow-Teller EC transition $3/2^- \to 5/2^-$. 
   %\textbf{ | Normalized partial half-lives for the Gamow-Teller EC transition $3/2^- \to 5/2^-$.}
   The first line lists the atomic orbitals $x$ with a positive EC $Q$ value, the second line shows the corresponding electron binding energies of the daughter $^{159}$Tb$^*$ atom~\cite{X-Ray_Data_Booklet}. The last line lists the resulting partial EC half-lives after normalizing to the total half-life 2.08$\times$10$^5$ years of the Gamow-Teller transition from \cite{NSR1968MY01}. The level P1 is calculated using the G{\lowercase{\scshape RASP}}2018 software package ~\cite{GRASP2018} for an isolated atom of $^{159}$Tb$^*$ in the configuration [Xe](4f)$^{10}$(6s)$^1$.
   }
\begin{ruledtabular}
\begin{tabular*}{\textwidth}{cccccc}
$x$ &N1 &N2&O1&O2 & P1
%\footnote{\label{fn:b} Calculated using the G{\lowercase{\scshape RASP}}2018 software package ~\cite{GRASP2018} for an isolated atom of $^{159}$Tb$^*$ in the configuration [Xe]$4f^{10}6s^1$.} 
\\ 
\hline
$\varepsilon_x$~[eV] & 396& 322.4 & 45.6 & 28.7 & 9.5 \\
%$\lambda_x$~[$m_e$]&  6.2$\times$10$^{-33}$  &3.1$\times$10$^{-34}$ &2.0$\times$10$^{-33}$  &7.0 %$\times$10$^{-35}$ &1.4$\times$10$^{-34}$\\
%$\lambda_x$~[year$^{-1}$]&  1.5$\times$10$^{-4}$  &7.5$\times$10$^{-6}$ &4.9$\times$10$^{-5}$  &1.7 $\times$10$^{-6}$ &3.5$\times$10$^{-6}$\\
$t_{1/2}$~[year]&3.0$\times$10$^{5}$& 5.8$\times$10$^{6}$ & 8.9$\times$10$^{5}$ &2.6$\times$10$^{7}$ &1.3$\times$10$^{7}$  \\
   \end{tabular*}
   \label{table:Q-value-halflife}
   \end{ruledtabular}
\end{table}
%%%%%%%%%%%%%%%%%%%%%%%970.83(64)
%-----------------------------Table 3+4  --------------------------------
%
% 
%

%\textbf{Partial decay constants and EC spectrum.}
%To estimate the partial half-lives for the transitions to the excited states and the distribution of energy released in the EC process, we have run atomic Dirac-Hartree-Fock many-body calculations. 
The energy distribution of EC events is represented as the incoherent sum of the contributions of individual orbitals:
\begin{equation}
%\frac{\mathrm{d}\lambda(E)}{\mathrm{d}E}
\rho(E) = \frac{G_{\beta}^{2}}{(2\pi)^{2}}\sum_{x}n_{x}%
\mathcal{B}_{x}\beta_{x}^{2}C_{x} p_{\nu}(E_{\nu})E_{\nu} \frac{ \Gamma_{x}%
/(2\pi)}{(E-\varepsilon_{x})^{2}+\Gamma_{x}^{2}/4},\label{energy distribution}%
\end{equation}
where $E = Q_{\mathrm{EC}}^{i}-E_{\nu}$ 
%is the energy deposited in detector
, 
$Q_{\mathrm{EC}}^{i}$ is the $Q$ value of the decay,
$E_{\nu}$ is the neutrino energy, 
$\lambda(E)$ is the total decay probability in the interval $(E, Q_{\mathrm{EC}}^{i}-m_{\nu})$; 
$G_{\beta} = G_{\mathrm{F}}\cos\theta_{\mathrm{C}}$,
$G_{\mathrm{F}}$ is the Fermi constant and $\theta_{\mathrm{C}}$ is the Cabibbo angle;
$p_{\nu}(E_{\nu}) = \sqrt{E_{\nu}^2 - m_{\nu}^2}$ is the neutrino momentum, $\varepsilon_{x}$ is the energy of the electron hole with quantum numbers $x=(n,l,j)$ of the daughter atom, and $n_{x}$ is the occupation fraction of electrons in a partially filled shell $x$ of the parent atom ($n_{x}=1$ for closed shells). The shape factor $C_{x}$ contains the nuclear-structure information in terms of nuclear form factors \cite{Behrens1982}. $\Gamma_{x}$ is the intrinsic linewidth of the Breit-Wigner resonance centered at the energies $\varepsilon_{x}$.
The amplitudes $\beta_{x}$, which characterize the electron wave functions inside the nucleus, and the exchange-and-overlap factors $\mathcal{B}_{x}$ are given for a broad set of atomic numbers and orbitals, e.g., in \cite{Bambynek1977} and here calculated for all orbitals of $^{159}$Dy and $^{159}$Tb$^{\ast}$ atoms by using the atomic structure software package G{\lowercase{\scshape RASP}}2018 \cite{GRASP2018}. The nuclear charge density is given by the Fermi distribution with the root mean square radius of
$R_{\mathrm{nucl.}}=5.1$ fm and thickness $2.3$ fm. The parent $^{159}$Dy atom is in the ground state, while the daughter atom $^{159}$Tb$^{\ast}$ is described by the electron wave functions depending on the hole $x$.
Electrons of the daughter atom inherit quantum numbers from the configuration [Xe](4f)$^{10}$(6s)$^{2}$ of the parent $^{159}$Dy atom. The exchange-and-overlap factors $\mathcal{B}_{x}$ calculated in the Vatai approach \cite{Bambynek1977} deviate from unity by 25\% or less.

The total decay constant $\lambda \equiv \lambda (Q_{\mathrm{EC}}^i-m_{\nu})$ is calculated from
\begin{equation}
\lambda(E) = 
\int_{0}^{E}
\rho(E^{\prime})dE^{\prime}.
\end{equation}
In the narrow-width approximation $\lambda \approx\sum_{x}\lambda_{x}$; the partial decay constants equal
\begin{equation}
\label{eq:lambda}
\lambda_{x}=\frac{G_{\beta}^{2}}{(2\pi)^{2}} n_{x}\mathcal{B}_{x}\beta_{x}%
^{2}C_{x}p_{\nu}(Q_{\mathrm{EC}}^{i}-\varepsilon_{x})(Q_{\mathrm{EC}}^{i}-\varepsilon_{x}).
\end{equation}

%For the presently discussed transitions to the $5/2^{-}$ state (an allowed Gamow-Teller transition) and $11/2^{+}$ state (a third-forbidden unique transition) the shape factor contains only one nuclear form factor in the leading order. 

For the presently discussed transitions to the $5/2^{-}$ state and $11/2^{+}$ state the shape factor contains only one nuclear form factor in the leading order. 

For the EC to the $5/2^-$ state the shape factor can be written as
%\begin{equation}
%\label{eq:GT-NME}
$C_x = (^{\mathrm{A}}F^{(0)}_{101})^2$,
%\end{equation}
with the nuclear form factor given in terms of the Gamow-Teller nuclear matrix element as
\begin{equation}
\label{eq:GT-NME2}
^{\mathrm{A}}F^{(0)}_{101} = -\frac{g_{\rm A}}{\sqrt{2J_i+1}}M_{\rm GT}.
\end{equation}
Here $g_{\rm A}$ is the strength of the weak axial coupling, $J_i$ the angular momentum of the initial state, and $M_{\rm GT}$ the Gamow-Teller nuclear matrix element \cite{JSuhonen2007}. In fact, for this decay transition we do not need the value of the form factor $^{\mathrm{A}}F^{(0)}_{101}$ since {we normalize $\lambda$} by the available half-life for the Gamow-Teller transition, derived from the measured branching~\cite{NSR1968MY01} and the total half-life~\cite{NNDC}. 
For this transition the experimental binding energies
%, computed decay constants 
and normalized partial half-lives are listed in Table~\ref{table:Q-value-halflife}.

The summation in Eq.~(\ref{energy distribution}) runs over the electron orbitals shown in  Table~\ref{table:Q-value-halflife}, as well as over the M1 and M2 orbitals. %with $\varepsilon_{x} > Q_{\mathrm{EC}}^{i}$. 
Although M1 and M2 are outside the kinematically accessible energy region, the tails of their Breit-Wigner amplitudes have a significant effect on the number of events for $E \lesssim Q_{\mathrm{EC}}^i$. 
The electromagnetic decay widths $\Gamma_{x}$ of the N1, N2, M1 and M2 electron holes in $^{159}$Tb$^*$ atom are taken from Ref.~\cite{CAMPBELL20011}; the data for $x=$ O1, O2, P1 are not available, so we assume $\Gamma_{\mathrm{O1,O2,P1}}=\Gamma_{\mathrm{N2}} = 5.26$ eV.
The widths of the levels N1, N2, M1 and M2 closest to the threshold are known with an accuracy of 10\%, 10-15\%, 5\% and 5-10\%, respectively~\cite{CAMPBELL20011}. The corresponding uncertainties in the spectrum do not exceed 30\%, while the integral over the spectrum is almost independent of the level widths. The experimental error in $Q_{\mathrm{EC}}^i$ introduces through the phase space volume about 50\% uncertainty in the half-life estimates.

\begin{figure}[!htpb]
\begin{center}
\includegraphics[angle = 0,width=0.45\textwidth]{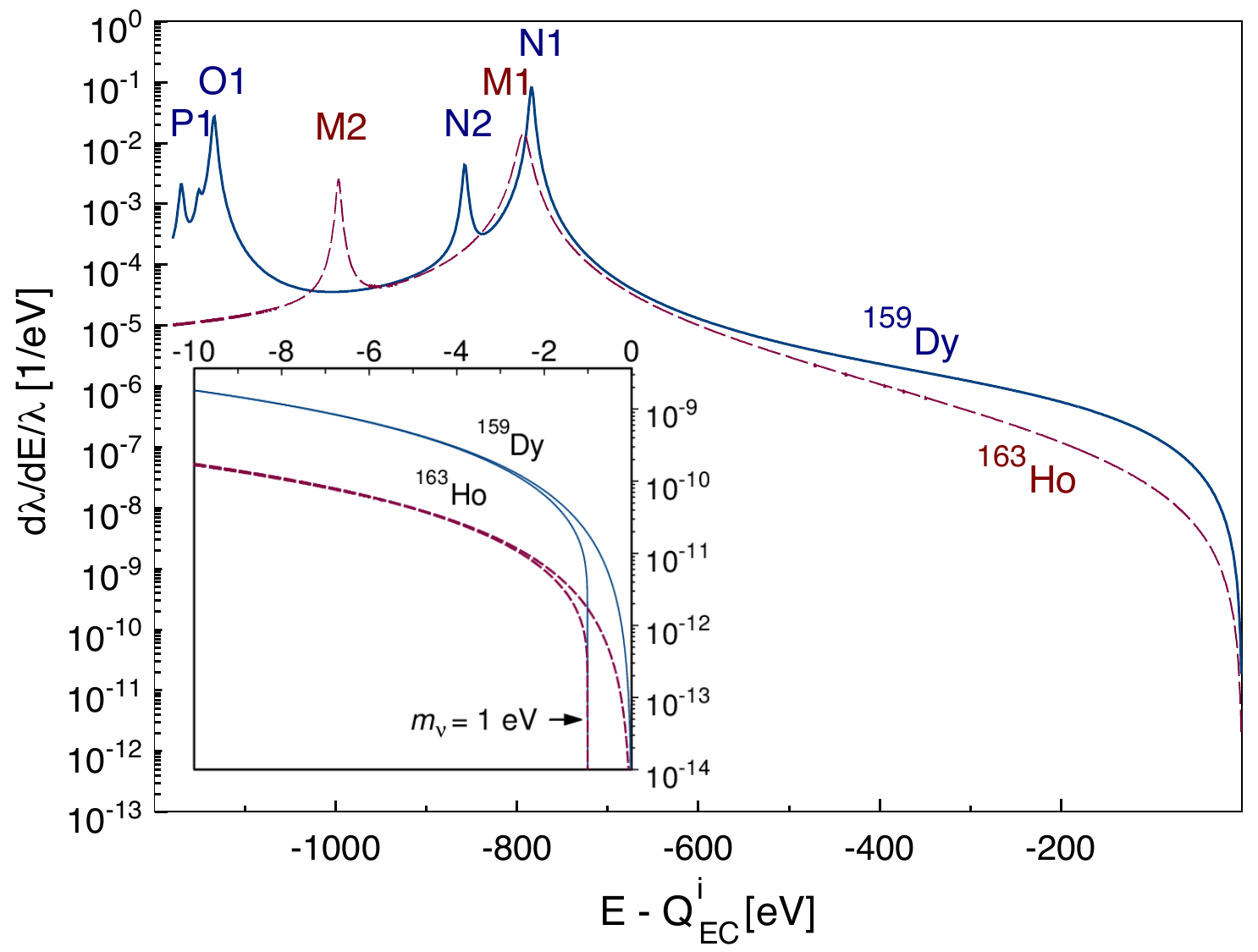}
\end{center}
\vspace{-6mm}
\caption{(Color online)
The solid curve describes the normalized distribution over the released energy of EC events in the $^{159}$Dy atom with the transition to the $^{159}$Tb atom having a nucleus in the $5/2^-$ excited state.
%The solid curve describes the normalized distribution of the released energy in the EC process with the $^{159}$Dy atom, leading to the transition to the $^{159}$Tb atom with the nucleus in the $5/2^-$ excited state.
N1, N2, O1 and P1 indicate electron holes of the $^{159}$Tb atom;
the O2 hole is barely discernible and not labeled. The dashed curve shows the normalized distribution in energy of the 
$^{163}$Ho EC events. M1 and M2 are electron holes of the $^{163}$Dy atom. The energy $E$ released in the electron capture takes values over the entire kinematically allowed region of the
$^{159}$Dy decay.
$Q_{\mathrm{EC}}^i$ is the difference in energy of the parent and daughter atoms. A larger fraction of events lands near the endpoint for $^{159}$Dy than for $^{163}$Ho. The inset in the lower left part of the figure shows on an enlarged scale the energy spectra of dysprosium and holmium close to the threshold value to illustrate the effect of the neutrino masses of 1 and 0~eV$/c^2$.
}
\label{fig10}%
\end{figure}

The 
%\textcolor{red}
{computed calorimetric $^{163}$Ho} spectrum of Fig.~\ref{fig10} takes the electron orbitals M1, M2, N1, N2, O1 and O2 into account with the parameters given in Ref.~\cite{Gastaldo2017}. The distances from the endpoint to the nearest peak for dysprosium (N1) and holmium (M2) are almost the same. Proximity of the M1 and M2 orbitals of dysprosium to the endpoint partly compensates the difference between the absolute EC rates of dysprosium and holmium at $E \lesssim Q_{\mathrm{EC}}^i$. The normalized cumulative distribution of the EC events near the endpoint equals $(\lambda - \lambda(E))/\lambda \approx C_{\nu} p_{\nu}^{3}(E_{\nu})$, where $C_{\nu}=0.0061$/keV$^{3}$ for dysprosium and $0.00056$/keV$^{3}$ for holmium. The M1 and M2 orbitals increase the number of events in the endpoint region by an order of magnitude.
%\textcolor{red}
{
The same absolute numbers of events near the endpoint are provided by the ratio 
between the numbers of dysprosium atoms decaying to Tb$^*(5/2^{-})$ and holmium atoms:
$R(^{159}\textrm{Dy}/^{163}\textrm{Ho}) = 
T_{1/2}(^{159}\mathrm{Dy}\rightarrow ^{159}\mathrm{Tb}^{\ast}(5/2^{-}))/T_{1/2}(^{163}\mathrm{Ho})C_{\nu }(^{163}\mathrm{Ho}%
)/C_{\nu }(^{159}\mathrm{Dy})=4.2$, 
while the total numbers of atoms before the filtering are in the ratio $R_0(^{159}\textrm{Dy}/^{163}\textrm{Ho}) = 4.2/1.9\times10^{-6} = 2.2\times10^6.$}
%\textcolor{red}
{The smallness of $C_{\nu }$ values limits, due to statistical requirements, the sensitivity of EC experiments measuring the mass of electron neutrino. To improve sensitivity, reliable parameterization of the energy spectrum away from peaks is also necessary, taking into account the dependence of the electron level widths and decay constants on energy.}
Decays accompanied by shake-up and shake-off excitations with the associated formation of multiple holes in the electron shell generate a fine structure of the spectrum~\cite{Robertson:2014fka,Faessler:2015txa,Faessler:2015pka,Faessler:2014xpa,Faessler:2016hxd,DeRujula:2016fdu}, which is experimentally visible in holmium EC and is described well theoretically~\cite{Brass:2017kov,Brab:2020uzx}.

%Neglecting them reduces the dysprosium coefficient $C_{\nu}$ to $0.00061$/keV$^{3}$.

%---------------------Fig. 4-----------------------
%---------------------Fig. 4-----------------------

%Both spectra are normalized to unity and it is clear that bigger fraction of events land near the endpoint for $^{159}$Dy than for $^{163}$Ho. 

The decay to the $11/2^+$ state gathers contributions from the M1-M5, N1-N7, O1-O3, and P1 atomic orbitals. 
The decay rate involves one nuclear form factor which we have computed using the microscopic interacting boson-fermion model (IBFM-2). 
%, see Methods). 
In this manner, we obtain an estimate of the half-life of $t_{1/2} \sim 10^{25}$ years for this transition, thus excluding it as a candidate for electron-neutrino mass measurements. There is also no experimental evidence for the existence of this transition.

The transition to the $5/2^-$ state has an experimentally measured half-life of 2.08$\times$10$^5$ years~\cite{NSR1968MY01}. This measured half-life can be used, together with the computed partial decay constants $\lambda_x$, to determine the normalized partial half-lives of the dominant EC channels. 
%These half-lives are given in   Table~\ref{table:Q-value-halflife}. 
%together with the computed $\lambda_x$. 
Using the computed decay constants and the IBFM-2 computed nuclear matrix element one obtains a theoretical half-life which is consistent with the measured one.
%Figure~\ref{fig10} shows distribution of the EC events in energy. 
Figure~\ref{fig10} shows the calculated EC spectrum.
%distribution. 
For comparison, the spectrum is also given for $^{163}$Ho. Both spectra are normalized to unity. It is clear that a larger fraction of events lands near the endpoint for $^{159}$Dy than for $^{163}$Ho. This is mostly due to M1 and M2 orbitals which, although energetically forbidden (see  Table~\ref{table:Q-value-AME}) for EC with $^{159}$Dy, affect the distribution due to the low energy tails of the M1 and M2 resonances in the endpoint region.
%(see Fig.~\ref{fig:level-sheme}).
%the proximity of these orbitals to the endpoint (see Fig.~\ref{fig:level-sheme}). 
%
% June 7, 2021. Tommi: The following commented out, I think it is repetition of what has been already said earlier.

%In conclusion, our findings reveal that the $Q_{\mathrm{EC}}$ of 1.18(19)~keV for the transition $^{159}$Dy(3/2$^-$) $\rightarrow$ $^{159}$Tb$^*$(5/2$^-$)  is  lower than presently running or planned direct neutrino mass experiments  using $^{163}$Ho, with the lowest ground-to-ground state $Q_{\mathrm{EC}}$.
In conclusion, our findings reveal that the $Q_{\mathrm{EC}}$ of 1.18(19)~keV for the transition $^{159}$Dy(3/2$^-$) $\rightarrow$ $^{159}$Tb$^*$(5/2$^-$) is lower than the ground-to-ground state $Q_\mathrm{EC}$ of $^{163}$Ho, which is utilized in presently running or planned direct neutrino mass experiments. 
%  using $^{163}$Ho, with the lowest ground-to-ground state $Q_{\mathrm{EC}}$.
%
Therefore, this allowed transition, with a universal spectral shape driven by a single decay matrix element and known branching ratio, becomes a potential candidate for effective electron neutrino mass measurements. Proximity of $Q_{\mathrm{EC}}$ and atomic lines N1, M1, and M2 with values of 0.79(19) keV, -0.78(19) keV, and -0.58(19) keV, respectively, indicates a significant potential of this EC transition
%as an important alternative to $^{163}$Ho 
for a self-calibrated and high-sensitivity EC experiment in the direct neutrino mass determination. 
%coincident registration of 363.5~keV and 305.5~keV de-excitation gamma-rays from the $5/2^-$ state of the nucleus. There is likely to be a 15.4~keV E1 transition available as well~\cite{NSR1968MY01}.
%\textcolor{red}
{The background from the EC to other states of $^{159}$Tb can be suppressed by coincident registration of de-excitation gamma-rays from the $5/2^-$ state of the nucleus. 
Such event selection is used in the search for neutrinoless double electron capture accompanied by nuclear excitations \cite{Blaum2020}. Decay to the $5/2^-$ level has a branching ratio of only $1.9\times10^{-6}$. In order to achieve sub-eV sensitivity, 
the measurement of the neutrino mass requires 
reliable coincidence measurements between the calorimeter and the $\gamma$ detector to identify only a very small fraction of total events, as well as a low background and a high counting rate of microcalorimeters.
}

We also want to point out that the Gamow-Teller EC transition to the $5/2^-$ state also serves as one of the most prospective transitions for a possible relic anti-neutrino capture experiment~\cite{lee2019new}. Here the very small $Q$ value, reported in this work, implies a promisingly high sensitivity to relic neutrinos requiring orders of magnitude less active material than needed for other suggested candidate nuclei like $^{163}$Ho or $^{157}$Tb.

We acknowledge the staff of the accelerator laboratory of University of Jyv\"askyl\"a (JYFL-ACCLAB) for providing stable online beam and J.~Jaatinen and R.~Sepp\"al\"a for preparing the production target. We thank the support by the Academy of Finland under the Finnish Centre of Excellence Programme 2012-2017 (Nuclear and Accelerator Based Physics Research at JYFL) and projects No. 306980, 312544, 275389, 284516, 295207, 314733, 318043, 327629 and 320062. The support by the EU Horizon 2020 research and innovation program under grant No. 771036 (ERC CoG MAIDEN) is acknowledged.

\bibliographystyle{apsrev4-1}
\bibliography{my-final-bib-from-jabref_titles}

\end{document}